\documentclass[aps,pra,twocolumn,showpacs,groupedaddress]{revtex4}
\newcommand{\ket}[1]{|#1\rangle}
\newcommand{\bra}[1]{\langle #1|}
\newcommand{\Tr}{\text{Tr}}
\usepackage{graphicx}
\begin{document}
\title{Entanglement of Fermi gases in a harmonic trap}
\author{X.X. Yi}
\affiliation{Department of physics, Dalian University of
Technology, Dalian 116024, China}
\date{\today}
\begin{abstract}
For both cases with and without interactions, bipartite
entanglement of two-fermions from a Fermi gas in a trap is
investigated. We show how the entanglement  depends on the
locations of the two fermions and the total number of particles.
Fermions at the verge of trap have longer entanglement distance
(beyond it, the entanglement disappears) than those in the center.
We derive a lower limitation to the average overlapping for two
entangled fermions in the  BCS ground state, it is shown to be
$\sqrt{Q/2M}$, a function of Cooper pair number $Q$ and total
number of occupied energy levels $M$.
\end{abstract}
 \maketitle

As one of the most characteristic features of quantum systems,
quantum entanglement lies at the heart of the difference between
the quantum and classical multi-particle world. It is the
phenomenon that enables quantum information processing and quantum
computing\cite{bouwmeester00, nielsen00}. Quantum entanglement is
usually considered as existing between different degrees of
freedom of two or more particles with mutual interactions, it is
only recently that researchers have started to investigate
entanglement in systems containing a large number of particles, in
particular in a noninteracting Fermi gas\cite{vedral03, oh04,
lunkes05}. Entanglement seems to play a crucial role in condensed
matter systems, and has shown of relevance to thermodynamical
quantities such as the degeneracy  pressure\cite{lunkes04} and the
number density of the gas\cite{lunkes05}, multipartite
entanglement is also promising to make breakthrough in solving
unsolved problems such as high-$T_c$
superconductivity\cite{vedral04}.

Entanglement for noninteracting Fermi gases in a free space has
already been studied\cite{vedral03, oh04, lunkes05}. It was found
that all entanglement vanishes if the relative distance
$|r-r^{\prime}|$ between electrons is greater than the
entanglement distance $1/k_F$, where $k_F$ denotes the Fermi
momentum. In this situation, quantum entanglement is purely due to
Fermi statistics and not due to any physical interaction. A
natural question then arises, is this a general property of
noninteracting Fermi gases? or it is special for the Fermi gas in
a free space, and how the interactions among the fermions
influence the entanglement?

In this paper, we will try to answer this question by studying the
entanglement in noninteracting Fermi gases trapped in a harmonic
trap, and examining  the effect of inter-fermion interactions on
the entanglement. As you will see, bipartite entanglement measured
by Wootters concurrence  depends not only on the relative distance
between two fermions  but also on the total number of particles,
the larger the number of fermion $N$, the shorter the entanglement
distance. The  entanglement distance is no longer a constant
$1/k_F$ as in the free space,  it is related to locations of the
two fermions. Our numerical simulations show that the entanglement
distance is longer at the verge more than that at the center of
the trap. Further, we show the effect of interactions on the
bipartite entanglement for the Fermi gas. The model adopted is the
reduced BCS Hamiltonian, it is  shown that the bipartite
entanglement of BCS ground state sharply depends on the
overlapping of the time-reversed states, the minimum average
overlapping is $\sqrt{Q/2M}$ depending on the ratio of Cooper pair
number $Q$ to the total number of occupied energy levels $M$, this
indicates bipartite entanglement in the conventional BCS ground
state. These results clearly establish the fact that entanglement
should be taken into account when studying macroscopic observable,
even if the system is in its ground state and the constituents of
the system are noninteracting as the fermions in a trap.

Suppose we have a collection of noninteracting Fermi gas in
harmonic traps. The Fermi gas may be electrons in a metal or
ultracold atoms. Unless stated otherwise, we treat the Fermi gas
in this paper as  ultracold atoms, but the representation is
applicable for all Fermi gases. At zero temperature, the atom gas
is in its lowest energy configuration. All energy states are
occupied up to the level $M=N/2$, where $N$ is the total number of
atom and assumed to be even (the case with odd number of particles
will be discussed at the end of this paper). For simplicity, we
consider throughout this paper a one-dimensional harmonic trap.
This condition is met if the trapping frequencies in the other
directions are considerable large. The ground state of this system
is
\begin{equation}
\ket{\Psi_0}=\prod_{n=1}^{M}
b^{\dagger}_{n\uparrow}b^{\dagger}_{n\downarrow}\ket
{Vac},\label{gs1}
\end{equation}
where $\ket {Vac}$ denotes the vacuum, and $b^{\dagger}_{n\sigma}
(\sigma=\uparrow,\downarrow)$ creates an atom in state $\phi_n(x)$
with spin $\sigma$. We are interested in entanglement between two
atom spins at different locations. The solution to this problem
would answer the following question, suppose that one atom has
spin up at $x$, can we infer the direction of the spin of the
other atom at $x^{\prime}$? In order to answer this question, the
density matrix describing the spin state of two atoms at locations
$x$ and $x^{\prime}$ is needed. It can be defined
by\cite{vedral03}
\begin{equation}
\rho_{ss^{\prime};tt^{\prime}}=\bra{\Psi_0}\psi^{\dagger}_{t^{\prime}}(x^{\prime})\psi^{\dagger}_{t}(x)
\psi_{s^{\prime}}(x^{\prime})\psi_{s}(x)\ket{\Psi_0},
\end{equation}
where $\psi^{\dagger}_{t}(x)$ creates an atom of spin $t$ at the
location $x$. This density matrix also may be calculated by
\begin{equation}
\rho_{ss^{\prime};tt^{\prime}}=\Tr{(\ket{\Psi_0}\bra{\Psi_0}\cdot\ket{st(x)}
 \bra{s^{\prime}t^{\prime}(x^{\prime})}}),
\end{equation}
with $\ket{st(r)}$ standing for two-atom state with spins $s$ and
$t$ at the location $r$. Writing $\psi_s(t)$ in terms of the
annihilation operator $b_{n\sigma}$ and the  eigenfunctions of
harmonic oscillator $\phi_n(x)$, we obtain the density matrix in
the following form,
\begin{equation}
\rho_{ss^{\prime};tt^{\prime}}=N(x)N(x^{\prime})\delta_{ts}\delta_{t^{\prime}s^{\prime}}
-\delta_{ts^{\prime}}\delta_{t^{\prime}s}F^2(x,x^{\prime}),
\end{equation}
where
$F(x,x^{\prime})=\sum_{\alpha}^{M}\phi^*_{\alpha}(x^{\prime})\phi_{\alpha}(x)=F$,
$N(x)=\sum_{\alpha}|\phi_{\alpha}(x)|^2=N_x$. Term $F$ represents
a sum over overlapping of the two atoms at locations $x$ and
$x^{\prime}$, respectively. $N(x)$ is the number density of atom
at the location $x$. In basis $\{
\ket{\uparrow\uparrow},\ket{\uparrow\downarrow},
\ket{\downarrow\uparrow},\ket{\downarrow\downarrow}\},$ the
density matrix takes the form,
\begin{widetext}
\begin{equation}
\rho_{12}(x,x^{\prime})=\frac{1}{4N_xN_{x^{\prime}}-2F^2}\left(
\matrix{ N_xN_{x^{\prime}}-F^2 & 0 &0&0 \cr
 0 & N_xN_{x^{\prime}} & -F^2 &0  \cr
0  & -F^2 & N_xN_{x^{\prime}} &0  \cr 0 & 0& 0&
N_xN_{x^{\prime}}-F^2} \right),\label{dm1}
\end{equation}
\end{widetext}
where the subscript $1,2$ indicates that there are two atoms. The
entanglement of formation\cite{wootters97} measured by Wootters
concurrence can be  given by
\begin{equation}
C_{12}(x,x^{\prime})=\frac{2}{|4N_xN_{x^{\prime}}-2F^2|}\mbox{max}\{2F^2-N_xN_{x^{\prime}},0\}.
\end{equation}
Obviously, the Wootters concurrence $C_{12}$ is maximal when
$x=x^{\prime}$ and it equals to 1. The corresponding entanglement
state is the spin singlet
$1/\sqrt{2}(\ket{\uparrow\downarrow}-\ket{\downarrow\uparrow})$.
With the relative distance between the two atoms increasing, $F^2$
behaves as a damping function of $|x-x^{\prime}|$. As a
consequence, the entanglement decay with $|x-x^{\prime}|$
increasing, this was shown in figure 1,
\begin{figure}
\includegraphics*[width=0.8\columnwidth,
height=0.6\columnwidth]{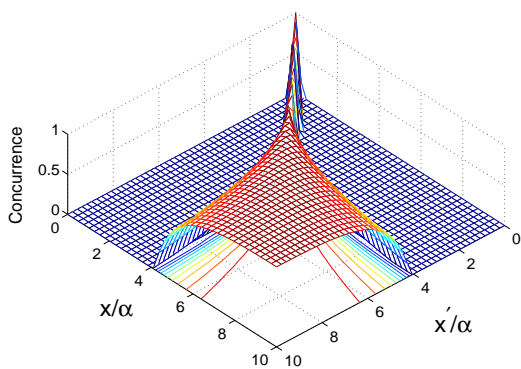} \caption{Wootters concurrence
of two atoms located at $x$ and $x^{\prime}$, respectively. The
figure was plotted for 20 trapped atoms.
$\alpha=\sqrt{m\omega/\hbar}$, $m$ is the mass of atom, and
$\omega$ is the trapping frequency.} \label{fig1}
\end{figure}
where the bipartite entanglement in a system with 20 atoms was
plotted as a function of locations $x$ and $x^{\prime}$ of the two
atoms. The entanglement arrives at the maximum value 1 at points
$|x-x^{\prime}|=0$ and decays  with atom separation increasing
from the pauli exclusion principle. This can be understood as a
result of more and more triplet states  mixed in with the singlet.
In contrast with the entanglement between two fermions in a free
space, the entanglement between the two in a harmonic trap appears
to be location dependent. It is clear from figure 1 that the
entanglement distance is longer for atoms at the verge of the trap
than that at the center. The reason for this is the following. The
bipartite entanglement roughly depends on how and how many triplet
state are mixed in with the singlet. At points $|x-x^{\prime}|=0$,
there are no triplet states involved in from the pauli exclusion,
so the bipartite entanglement is maximal at these points. With the
atom separation increasing, more triplet states are involved in,
the two atom state is then a weighted sum over the singlet and
triplet states, the weights depend on the locations of atoms.
Because the center of the trap is the favorite location for
particles to occupy, the weights at the verge benefit the
entanglement. The bipartite entanglement also depends on the total
number of atom, as shown in figure 2, where one atom was located
at $0.5\alpha$, $\alpha=\sqrt{m\omega/\hbar}$, $m$ is the mass of
atom, and $\omega$ stands for the trapping frequency. Clearly, the
larger the number of atom, the faster the damping of the bipartite
entanglement.
\begin{figure}
\includegraphics*[width=0.8\columnwidth,
height=0.6\columnwidth]{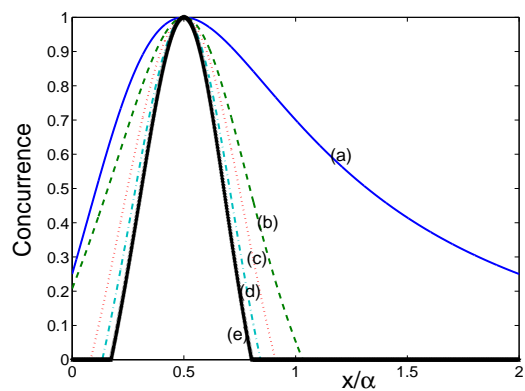} \caption{The bipartite
entanglement as a function of the second atom's location $x$.
(a)-(e) are for different total number of atoms. The number
corresponding to (a),...,(e) is 2,4,10,14,18, respectively.}
\label{fig2}
\end{figure}

Up to now, we have not considered the interaction between the
atoms. In the following, we study how the interaction influences
the entanglement of the two atoms. For repulsive interactions,
perturbation theory tells us that weak interactions modify the
eigenvalues and the corresponding eigenfunctions of the free
Hamiltonian, this results in a level shift to every eigenstate of
the free Hamiltonian, and consequently there are more
eigenfunctions participating in the summations in $F$, $N_x$, and
$N_{x^{\prime}}$. So, the damping in the dependence of
entanglement on the relative distance would faster than that
without interactions. For attractive interaction, we will
investigate the entanglement in trpped atoms by using the reduced
BCS model. The reduced BCS Hamiltonian has received much attention
as a result of  effort to understand pairing correlations in
nanoscale metallic system \cite{dukelsky00, ralph01}. The model
Hamiltonian is\cite{delft96},
\begin{equation}
H_{BCS}=\sum_{j=1}^{M} \varepsilon_j n_j-d\lambda\sum_{j\neq
k}^{M} b^{\dagger}_{j+}b^{\dagger}_{j-}b_{k+}b_{k-},
\end{equation}
where $j,k=1,...,M$ labels a set of doubly degenerate single
particle energy levels with energies $\varepsilon_j$, $\lambda$ is
the dimensionless coupling, and $d$ is the mean level spacing.
$b_{j\pm} (b^{\dagger}_{j\pm})$ represent the annihilation
(creation) operators for atoms at level $j$ with the labels $\pm$
referring to pairs of time-reversed states, and $n_j$ was defined
as $n_j=b^{\dagger}_{j+}b_{j+}+b^{\dagger}_{j-}b_{j-}$, the atom
number operator for level $j$. The ground state entanglement
(called ALC - average local concurrence) share among different
energy levels in this model was investigated in \cite{dunning05},
it shown a simple relation between the ALC and the order
parameter. The conventional BCS theory employs a grand canonical
ensemble, by the Bogoliubov transformation
$\gamma_{j1}=u_jb_{j+}-v_jb^{\dagger}_{j-},$
$\gamma_{j0}=u_jb_{j-}+v_jb^{\dagger}_{j+},$ with $u_j^2+v_j^2=1$,
$4u_j^2v_j^2=\Delta^2/(\varepsilon_j^2+\Delta^2)$, and
$\Delta=\lambda d\sum_{j=1}^{M}\langle b_{j-}b_{j+}\rangle,$ it
gives
\begin{eqnarray}
H_{BCS}&=&\sum_j[\varepsilon_j(u_j^2-v_j^2)-2\Delta
u_jv_j](\gamma^{\dagger}_{j1}\gamma_{j1}+
\gamma^{\dagger}_{j0}\gamma_{j0})\nonumber\\
&+&\mbox{constant}.\label{bcs}
\end{eqnarray}
Instead considering entanglement between two atom spins, we here
consider entanglement between two atoms in the two time-reversed
states $+$ and $-$. The density matrix representing this entangled
state can be defined as
\begin{equation}
\rho_{ss^{\prime};tt^{\prime}}^{BCS}=\Tr{(\ket{BCS}\bra{BCS}\cdot\ket{st(x)}
 \bra{s^{\prime}t^{\prime}(x^{\prime})})},\label{density2}
\end{equation}
where $\ket{BCS}$ is the well-known ground state in the BCS model,
$\ket{BCS}=\prod_{j=1}^{M}(u_j+v_jb^{\dagger}_{j+}b^{\dagger}_{j-})\ket
0,$ $s,t,s^{\prime}, t^{\prime}=+,-.$ By the standard precedure,
Eq. (\ref{density2}) yields,
\begin{equation}
\rho_{ss^{\prime};tt^{\prime}}^{BCS}=(\sum_j|v_j|^2)^2\delta_{ts}\delta_{t^{\prime}s^{\prime}}
-\delta_{ts^{\prime}}\delta_{t^{\prime}s}\mbox{Re}(f(x,x^{\prime})v^2(x,x^{\prime})),
\end{equation}
where
\begin{eqnarray}
f(x,x^{\prime})&=&\sum_j\phi_j(x)\phi_j^*(x^{\prime})=f,\nonumber\\
v^2(x,x^{\prime})&=&\sum_j v_j^2\phi_j^*(x)\phi_j(x^{\prime})=v^2.
\end{eqnarray}
Here, Re$(...)$ denotes the real part of $(...)$. In basis  $\{
\ket{++},\ket{+-}, \ket{-+},\ket{--}\},$ the density matrix
defined in Eq.(\ref{density2}) follows,
\begin{widetext}
\begin{equation}
\rho_{12}^{BCS}(x,x^{\prime})=\frac{1}{4[\sum_jv_j^2]^2-2\mbox{Re}(fv^2)}\left(
\matrix{ [\sum_jv_j^2]^2-\mbox{Re}(fv^2) & 0 &0&0 \cr
 0 & [\sum_jv_j^2]^2 & -\mbox{Re}(fv^2) &0  \cr
0  & -\mbox{Re}(fv^2) & [\sum_jv_j^2]^2 &0  \cr 0 & 0& 0&
[\sum_jv_j^2]^2-\mbox{Re}(fv^2)} \right),\label{matrix2}
\end{equation}
\end{widetext}
$\sum_j v_j^2$ equals the total number $Q$ of Cooper pairs in
state $\ket{BCS}$, while $f$ characterize the overlapping of the
two particles. The state Eq.(\ref{matrix2}) is entangled iff the
Peres-Horodecki condition\cite{peres95} is met, i.e.,
$2\mbox{Re}(fv^2)-Q^2>0$. Assume
$\phi_j^*(x)\phi_j(x^{\prime})=y,$ $y$ is a constant, the
Peres-Horodecki condition leads to $|y|>\sqrt{Q/2M}$ $(<1/\sqrt
2)$. It is the restriction to the overage overlapping for
bipartite entanglement. With the assumption
$\phi_j^*(x)\phi_j(x^{\prime})=y,$ it is easy to write down the
concurrence
$C_{12}^{BCS}(x,x^{\prime})=\mbox{max}\{\frac{(2|y|^2-Q/M)}{(2Q/M-|y|^2)},0\}.$
The maximal entanglement $C_{12}^{BCS}(x,x^{\prime})=1$ is
obtained at $|y|^2=Q/M.$ For conventional BCS state, $\phi_j(x)$
may have the form of $e^{ip_jx}$, choose $x=x^{\prime}$,
$|y|=1>1/\sqrt 2.$ So two electrons in the conventional BCS state
are entangled as long as the separation $L$ is less than the
entanglement distance satisfying
$|\phi_j^*(x)\phi_j(x+L)|=|y|=\sqrt{Q/2M}$. However, it is  unsure
for a system with $\phi_i(x)$ taking the eigenstates of harmonic
oscillator. It is easy to show that
$|\phi_j^*(x)\phi_j(x^{\prime})|<1/\sqrt 2$ for any $j$, $x$ and
$x^{\prime}$, which indicates that small ratio of the Cooper pair
number to the total number of occupied levels might guarantee  the
entanglement existing  in such a system.

We now consider the case when the total number of particles is
odd. Intuitively, for a large number (say, $2N+1$) of atom, the
bipartite entanglement would behave like that with $2N$ particles.
This is the case indeed as you will see. Let us first analyze the
case without inter-particle interactions. Assume the $(2N+1)$th
atom is of spin up, the state of this $2N+1$ atoms may be written
as $b^{\dagger}_{M+1 \uparrow}\ket{\Psi_0}$, where $\ket{\Psi_0}$
was defined by Eq.(\ref{gs1}) for the $2N$ particles. Following
the calculation performed for Eq.(\ref{dm1}), we find that the
density matrix $\rho_{12}(x,x^{\prime})$ can be divided into two
matrices, the first represents contributions from the $2N$ atoms,
which takes the same form as in Eq.(\ref{dm1}), and the second is
a correction due to the $(2N+1)$th atom. The elements of the
second
 matrix $\sigma$ are,
\begin{eqnarray}
\sigma_{22}&=&2N_x|\phi_{M+1}(x^{\prime})|^2+2N_{x^{\prime}}|\phi_{M+1}(x)|^2,\nonumber\\
\sigma_{23}&=&-4F(x,x^{\prime})\phi_{M+1}^*(x)\phi_{M+1}(x^{\prime}),
\end{eqnarray}
 $\sigma_{11}=\sigma_{22}+\sigma_{23},$ $\sigma_{22}=\sigma_{33},$
$\sigma_{32}=\sigma^*_{23}$, and the others are zero. For a large
system ($N>>1$), $|\phi_{M+1}(x)|^2<<N_x$, and
$|\phi_{M+1}(x)\phi_{M+1}(x^{\prime})|<<|F(x,x^{\prime})|$.
Therefore, the correction $\sigma$ to the density matrix
$\rho_{12}(x,x^{\prime})$ can be neglected. However, this is not
the case if the system only consists of few atoms, the correction
due to the $(2N+1)$th atom should be taken into account to compute
the bipartite entanglement. Note that  exchanges of spin up with
spin down do not change $\ket{\Psi_0}$, but $b^{\dagger}_{M+1
\uparrow}\ket{\Psi_0}$. So matrix $\sigma$ is of relevance to the
spin of the $(2N+1)$th atom. But this does not affect the
bipartite entanglement under consideration, i.e., the bipartite
entanglement is independent of the spin of the $(2N+1)$th atom. In
the case of attractive interaction, the situation is similar if
the BCS ground state is simply
$b^{\dagger}_{M+1\uparrow}\ket{BCS}$. The situation becomes
complicated when $b^{\dagger}_{M+1\uparrow}\ket{BCS}$ is not the
ground state of the system\cite{braun97}. In order to calculate
the entanglement in BCS ground state, we have to find the ground
state first, it is beyond the scope of this paper.

Before concluding, it is worth mentioning that bipartite
(multipartite) entanglement as a properties between quantum
systems depends on definition of the two degrees (many degrees)
which share the entanglement. Some properties of entanglement in
the BCS model were studied in Ref.\cite{dunning05,dusuel05}, where
the entanglement was defined  among particles in different energy
levels\cite{dunning05}. The entanglement presented in this paper
is for two fermion spins at different locations $x$ and
$x^{\prime}$, it characterizes two-fermion's correlation at the
two locations.

In conclusion, we have shown that the bipartite entanglement in
noninteracting fermions trapped in harmonic traps depend on
particle number, relative distance and the locations of the two
fermions. The entanglement distance which characterize the maximum
separation of two entangled particles is longer at the verge of
the trap than that at the center; the larger the number of trapped
particles, the shorter the entanglement distance. For interacting
Fermi system, we have adopted the reduced BCS model to study the
entanglement in the BCS ground state. Reduced density matrix and
Wootters concurrence have been presented, the restriction on the
average overlapping $y$ has been derived  to be
$\sqrt{Q/2M}<|y|\leq \sqrt{Q/M}$, the lower limitation
$\sqrt{Q/2M}$ corresponds to concurrence zero, while the upper one
$\sqrt{Q/M}$ corresponds to maximal entangled states.

\vskip 0.3 cm We acknowledge financial support from NCET of M.O.E,
and NSF of China Project No. 10305002.

\end{document}